\documentclass{luxenote}

\title{LUXE: A new experiment to study non-perturbative QED in $e^-$-laser and $\gamma$-laser collisions.}
\author{Ruth Jacobs$^1$, on behalf of the LUXE collaboration}
\affil{$^1$Deutsches Elektronen-Synchrotron DESY, Notkestr. 85, 22607 Hamburg, Germany}
\date{\today}

\DocAuthors{A.~Person \\ B.~Person \\ C.~Person}
\DocEdBoard{D.~Person \\ E.~Person }
\LUXEDocNo{xxx-yyy-zzzz}
\draftversion{0.1}

\bibliography{references}

\begin{document}

\maketitle

\begin{abstract}
The LUXE experiment (Laser Und XFEL Experiment) is a new experiment in planning at DESY Hamburg using the electron beam of the European XFEL. At LUXE, the aim is to study collisions between a high-intensity optical laser and up to $16.5\,$GeV electrons from the Eu.XFEL electron beam, or, alternatively, high-energy secondary photons. The physics objectives of LUXE are to measure processes of Quantum Electrodynamics (QED) at the strong-field frontier, where QED is non-perturbative. This manifests itself in the creation of physical electron-positron pairs from the QED vacuum. LUXE intends to measure the positron production rate in a new physics regime at an unprecedented laser intensity. Additionally, the high-intensity Compton photon beam of LUXE can be used to search for physics beyond the Standard Model.
\end{abstract}

\vfill
\begin{center}
\textit{Presented at the 30th International Symposium on Lepton Photon Interactions at High Energies, hosted by the University of Manchester, 10-14 January 2022.}
\end{center} 
\clearpage





\section{Introduction}
The LUXE experiment intends to study Quantum Electrodynamics (QED) in a new, previously uncharted regime of strong electromagnetic background fields, where QED becomes non-perturbative. LUXE probes this new regime by studying collisions between the Eu.XFEL $16.5\,\text{GeV}$ electron beam and a $40\,\text{TW}$ optical laser, as well as collisions between a beam of secondary photons in the GeV energy regime and the laser. Due to the high laser intensity and the Lorentz boost of the initial electron beam, LUXE achieves field strengths close to, and above, the critical field stength, also known as the \textit{Schwinger limit} (in case of an electric field $E_{\text{cr}} = m^2_ec^3/(e\hbar)\approx1.32\times10^{18}\,\text{V/m}$)\footnote{Here, $m_e$ denotes the electron mass, $c$ the speed of light in vacuum, $e$ the electron charge and $\hbar$ the reduced Planck constant.}~\cite{Schwinger:1951}. Above the Schwinger limit, the work by the field over the Compton wavelength $\lambdabar=\hbar/(m_ec)$ of the probe electron is larger than the electron rest mass. As a result, the QED vacuum is polarized, which manifests itself in the creation of physical electron-positron pairs. LUXE aims to measure the positron production rate from high-energy photons impinging on the strong-field polarized vacuum, as well as study the onset of non-perturbativity at weak coupling in non-linear Compton scattering. For a recent review of strong-field QED processes and effects, see~\cite{Fedotov:2022ely}.\\

This article is structured as follows: Section \ref{sec:sfqed} gives an overview of the processes and parameters in strong-field QED, section \ref{sec:experiment} discusses the LUXE experimental setup and section \ref{sec:results} showcases the expected physics results of the experiment. The conclusion is presented in section \ref{sec:conclusion}.

\section{Strong-field QED}
\label{sec:sfqed}

Two parameters are typically used to describe the processes of non-perturbative QED in the regime studied by LUXE. The \textit{classical nonlinearity parameter} $\xi=m_ec^2E/(\hbar\omega_LE_\text{cr})$, where $\omega_L$ is the laser wavelength, $E$ is the electromagnetic field strength, and $E_\text{cr}$ is the Schwinger critical field,  characterizes the coupling between the high-intensity laser and the probe electron. If $\xi\sim\mathcal(O)(1)$, the QED perturbative expansion breaks down. The \textit{quantum non-linearity parameter} $\chi=eE\lambdabar/(mc^2)$ quantifies the field strength experienced by the probe electron in its rest frame in units of the Schwinger critical field, or the recoil of an electron emitting a photon. The two parameters are related via the equation $\chi=\xi\eta$, where $\eta=\gamma\hbar\omega_L(1+\cos\theta)/(m_ec^2)$ is the energy parameter, $\gamma$ is the relativistic Lorentz factor, and $\theta$ is the electron-laser crossing angle ($\theta\approx17.2^\circ$ in LUXE).\\

\begin{figure}[h]
     \centering
     \begin{subfigure}[b]{0.49\textwidth}
         \centering
         \includegraphics[width=\textwidth]{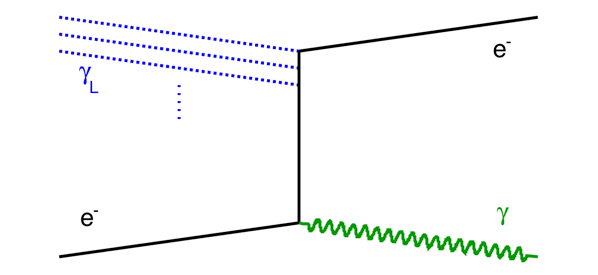}
         \caption{Non-linear Compton scattering.}
         \label{fig:feyncompton}
     \end{subfigure}
     \hfill
     \begin{subfigure}[b]{0.49\textwidth}
         \centering
         \includegraphics[width=\textwidth]{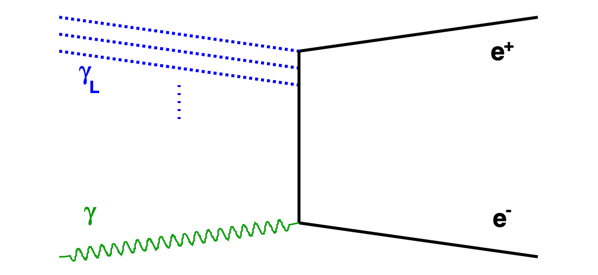}
         \caption{Non-linear Breit-Wheeler pair production.}
         \label{fig:feynbppp}
     \end{subfigure}
     \caption{Feynman diagrams of strong-field QED processes at LUXE~\cite{Abramowicz_2021}.}
\end{figure}

The processes indicating the transition from perturbative to non-perturbative QED are \textit{non-linear Compton scattering} (fig. \ref{fig:feyncompton}) and \textit{non-linear Breit-Wheeler pair production} (fig. \ref{fig:feynbppp})~\cite{PhysRevLett.26.1072}. In the case of non-linear Compton scattering, the probe electron absorbs multiple laser photons and emits a single high-energy Compton photon. For $\xi\gsim1$, the probability of the n-th order scattering process to occur, is proportional to $\xi^{2n}$. All orders in laser-electron coupling contribute to the non-linear Compton scattering amplitude. Experimentally, the non-linear Compton scattering manifests itself via a shift in the position of the kinematic Compton edge as a function of laser intensity $\xi$, as the electron acquires an effective mass $m_e^*=m_e\sqrt{1+\xi^2}$. It should be noted that the non-linear Compton edge shift occurs also in the classical limit of $\chi\ll1$. For $\chi\gsim1$, an additional, smaller ``quantum'' edge shift occurs, due to the modified electron recoil.\\

In the non-linear Breit-Wheeler process a real photon (produced either in non-linear Compton scattering in the LUXE $e^-$-laser mode, or via Bremsstrahlung in a target in the $\gamma$-laser mode), absorbs multiple laser photons and produces a physical electron-positron pair. This process does not have a classical equivalent in LUXE and can therefore only occur for $\chi\gsim1$. In the limit $\xi\gg1,\chi\ll1$, the Breit-Wheeler pair production rate scales as:
\begin{equation}
    \Gamma_\text{BW}\propto\exp\left[-\frac{3}{8}\frac{1}{\omega_\gamma(1+\cos\theta)}\frac{E_\text{cr}}{E_\text{L}}\right].
\end{equation}

LUXE aims to measure both non-linear Compton scattering and non-linear Breit-Wheeler pair creation as a test of strong-field QED and to extract the functional dependence on the parameters $\chi$ and $\xi$ to unprecedented precision. Furthermore, LUXE, in its $\gamma$-laser mode, will be the first experiment to probe non-linear Breit-Wheeler pair production with real high-energy photons.

\section{LUXE experimental setup}
\label{sec:experiment}

\begin{figure}[h]
     \centering
     \begin{subfigure}[b]{0.49\textwidth}
         \centering
         \includegraphics[width=\textwidth]{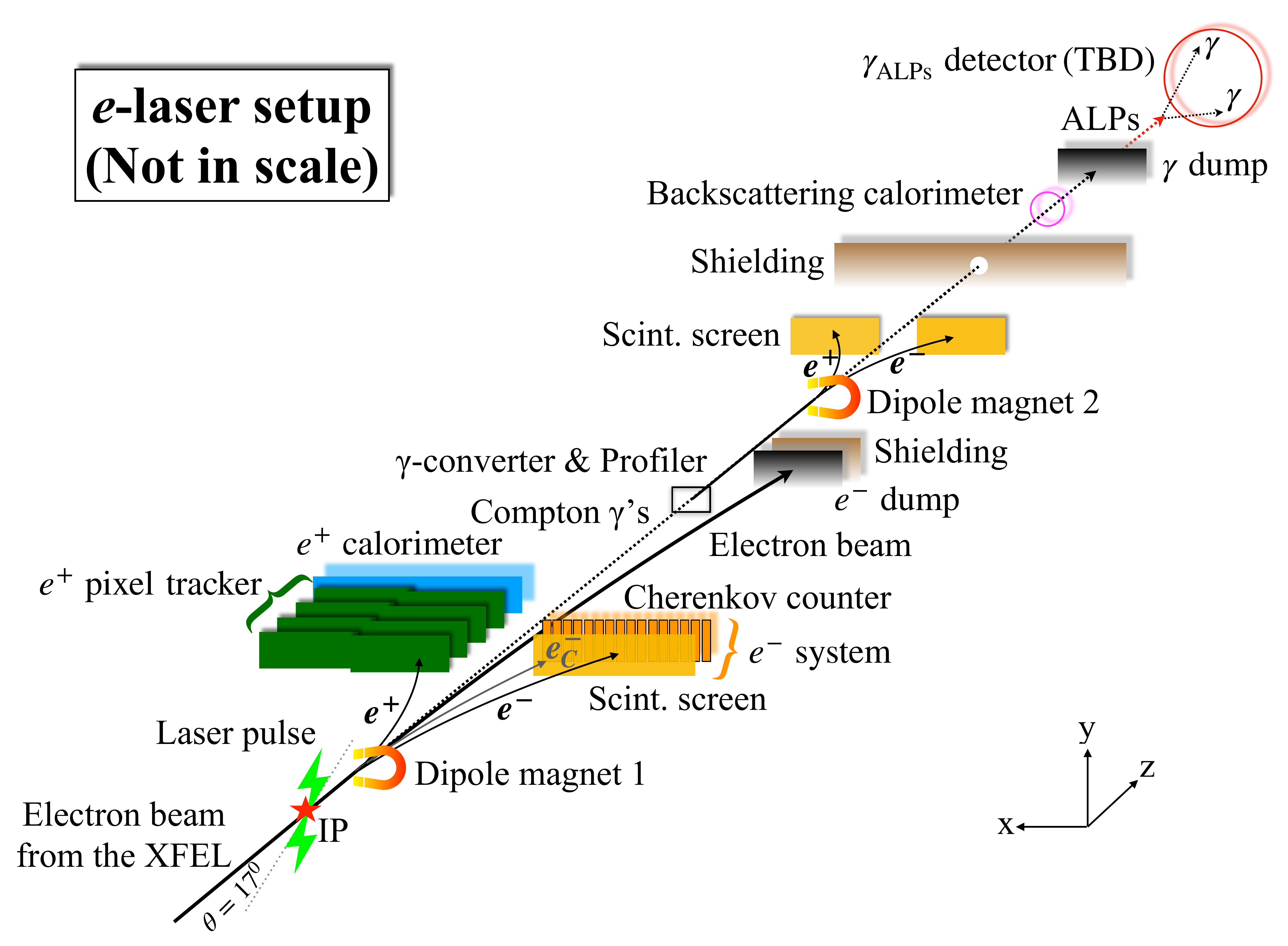}
         \caption{Electron-laser setup.}
         \label{fig:setuptrident}
     \end{subfigure}
     \hfill
     \begin{subfigure}[b]{0.49\textwidth}
         \centering
         \includegraphics[width=\textwidth]{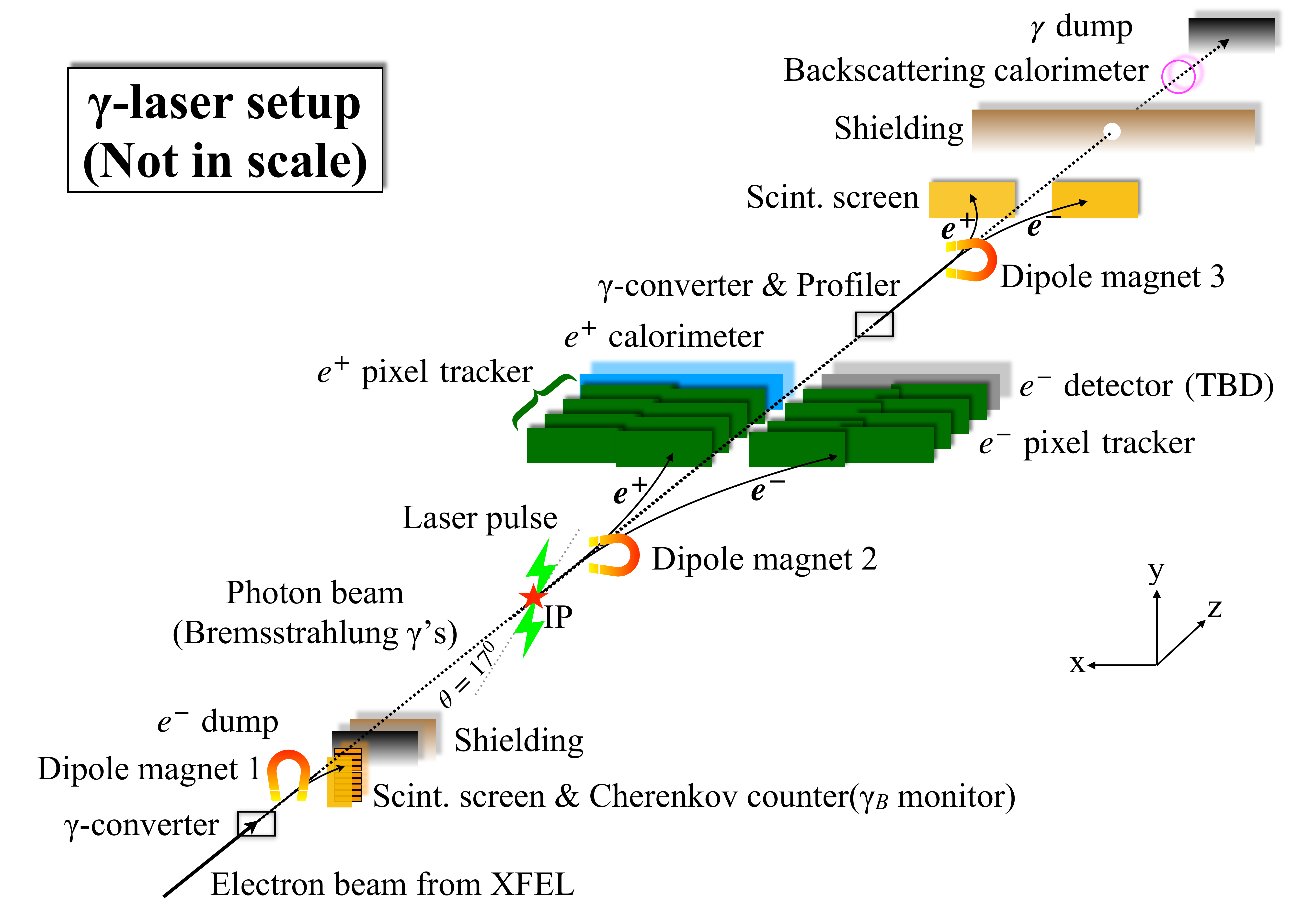}
         \caption{Photon-laser setup.}
         \label{fig:setupbppp}
     \end{subfigure}
     \caption{Sketch of the LUXE experimental setups~\cite{Abramowicz_2021}.}
      \label{fig:luxesetups}	
\end{figure}

A schematic sketch of the LUXE experimental setup for the $e^-$-laser and $\gamma$-laser setups is shown in figure \ref{fig:luxesetups}. The LUXE experiment will be situated at the end of the electron linear accelerator of the Eu.XFEL. A single bunch of the Eu.XFEL bunch train (2700 bunches in total) will be transported at a rate of $10\,\text{Hz}$ to the LUXE experimental area using a dedicated extraction beamline~\cite{beamlinecdr}.  In the $e^-$-laser mode the electron beam will collide directly with the high-intensity laser pulse. In the $\gamma$-laser mode, a converter target will be placed into the electron beam path, producing a high-energy photon beam via Bremsstrahlung. Alternatively the initial photon beam is produced by inverse Compton scattering (ICS) of the electron beam with a weaker laser pulse. The advantage of ICS production is the narrow photon energy bandwidth at the cost of a limited maximum photon energy, while the Bremsstrahlung production can reach photon energies of up to $16.5\,\text{GeV}$ with a broad energy distribution. In LUXE, the high-intensity laser pulse is triggered with a clock frequency of $1\,\text{Hz}$. Since the electron beam enters the experimental area at $10\,\text{Hz}$ repetition rate, the nine beam-only events in between collisions are used to study the background.

In the following sections, the core components of the LUXE experimental setup will be outlined in further detail.

\subsection{Optical laser}
The laser system foreseen for LUXE is an optical ($\lambda=800\,\text{nm}$) femto-second pulsed Titanium-Sapphire laser using the chirped-pulse-amplification technique~\cite{Strickland:1985gxr}. A staged approach is planned, using an upgradable laser system, which will deliver a power of $40\,\text{TW}$ in phase-0 ($\xi_\text{max}=7.9,\:\chi_\text{max}=1.5$) and subsequently will be upgraded to $350\,\text{TW}$ for phase-1 ($\xi_\text{max}=23.6,\:\chi_\text{max}=4.45$). During the data-taking in LUXE, $\xi$ and $\chi$ are varied by de-focusing and re-focusing the laser pulse at the interaction point. One of the main requirements of the LUXE laser is an exceptional shot-to-shot stability ($<1\%$ variation) and a precise measurement ($<5\%$ uncertainty) of the absolute laser intensity. To achieve this, a dedicated laser diagnostics suite using several complementary diagnostics methods on the interacting laser pulse will be employed. 

\subsection{Particle detectors}

One of the challenges of the LUXE experiment is the wide range of particle rates present in the different detector locations before and after the $e^-/\gamma$-laser interaction point (IP), and for the two different LUXE run modes. Table \ref{tab:luxerates} summarizes the particle rates for each location and run-mode.\\

\begin{table}
\begin{center}
\begin{tabular}{|c|c|c|} \hline
 \textbf{Location} & \textbf{$e^-$-laser particle rate} & \textbf{$\gamma$-laser particle rate} \\\hline
  initial target monitor & $-$ &  $10^5$  \\\hline
  post-IP $e^+$ & $10^{-4}-10^{4}$ & $10^{-4}-10^{1}$ \\\hline
  post-IP $e^-$ & $10^{3}-10^{8}$ & $10^{-4}-10^{1}$ \\\hline
  forward $\gamma$ & $10^{3}-10^{8}$ & $10^5$ \\\hline
\end{tabular}
\caption{Particle rates expected in LUXE for different locations in the experimental setup and run modes.}
\label{tab:luxerates}
\end{center}
\end{table}

The choice of detector technologies in each of the locations is motivated by their purpose as follows:

\begin{itemize}
    \item \textbf{Initial target area:} The aim of the initial target system is to monitor the energy spectrum of the high-energy photon beam produced via Bremsstrahlung or ICS in the $\gamma$-laser mode, by observing the energy spectrum of the electrons produced in the production process. The energy spectrum is resolved via a dipole spectrometer. Due to the high flux of electrons, a segmented air-filled \cer detector and a scintillator screen, read out by an optical camera, are used as detectors. In the $e^-$-laser case, the dipole is switched off and the electrons travel directly to the IP.
    \item \textbf{Post-IP area:} After the IP, electrons and positrons are deflected in opposite directions by a dipole magnet. In the $\gamma$-laser case, both electron and positron rates are comparatively low and a good background rejection is required. On the positron side, a 4-layer silicon tracking station is placed, followed by a high-granularity electromagnetic calorimeter, while a second silicon tracker is used on the electron side. In the $e^-$-laser case, the particle rates on the electron side are much higher, because of non-linear Compton scattering, therefore the silicon tracker is moved out of the detection plane and is replaced by a scintillator screen and camera system followed by a \cer detector, similar to the setup used in the initial target monitoring system.
    \item \textbf{Forward photon area:} The Compton photons created at the IP in the $e^-$-laser setup and the non-interacting part of the $\gamma$-laser initial photon beam traverse the LUXE setup undisturbed by the dipole spectrometers towards the forward part of the experiment, which is instrumented with three detectors. The gamma spectrometer system uses a thin target foil to convert part of the photon beam and uses a dipole spectrometer with a scintillator and camera system to reconstruct the photon energy. The gamma beam profiler uses a sapphire strip detector to precisely measure the spatial characteristics of the photon beam, allowing determination of the parameters $\xi$ and $\chi$ independently of the laser diagnostics suite. Finally, a lead-glass calorimeter measuring back-scattered particles from the photon beam dump is used to determine the total photon flux.
\end{itemize}

\subsection{Search for New Physics in LUXE}

In addition to probing strong-field QED physics processes, the enormous rate of photons produced in LUXE (mainly through non-linear Compton scattering in the $e^-$-laser setup) can be used to search for physics beyond the Standard Model (BSM). One possible scenario is axion-like particles (ALPs) being created via Primakoff production in the LUXE photon beam dump. By placing a calorimeter at a fixed distance behind the photon dump and ensuring a low background in this detection area, decays of the ALPs to two photons can be probed. With a $1\times1\,\text{m}^2$ detector, a sensitivity can be achieved that exceeds that of other existing experiments searching for ALPs~\cite{LUXENPOD}.

\section{Expected results}
\label{sec:results}

\begin{figure}[h]
     \centering
     \begin{subfigure}[b]{0.49\textwidth}
         \centering
         \includegraphics[width=\textwidth]{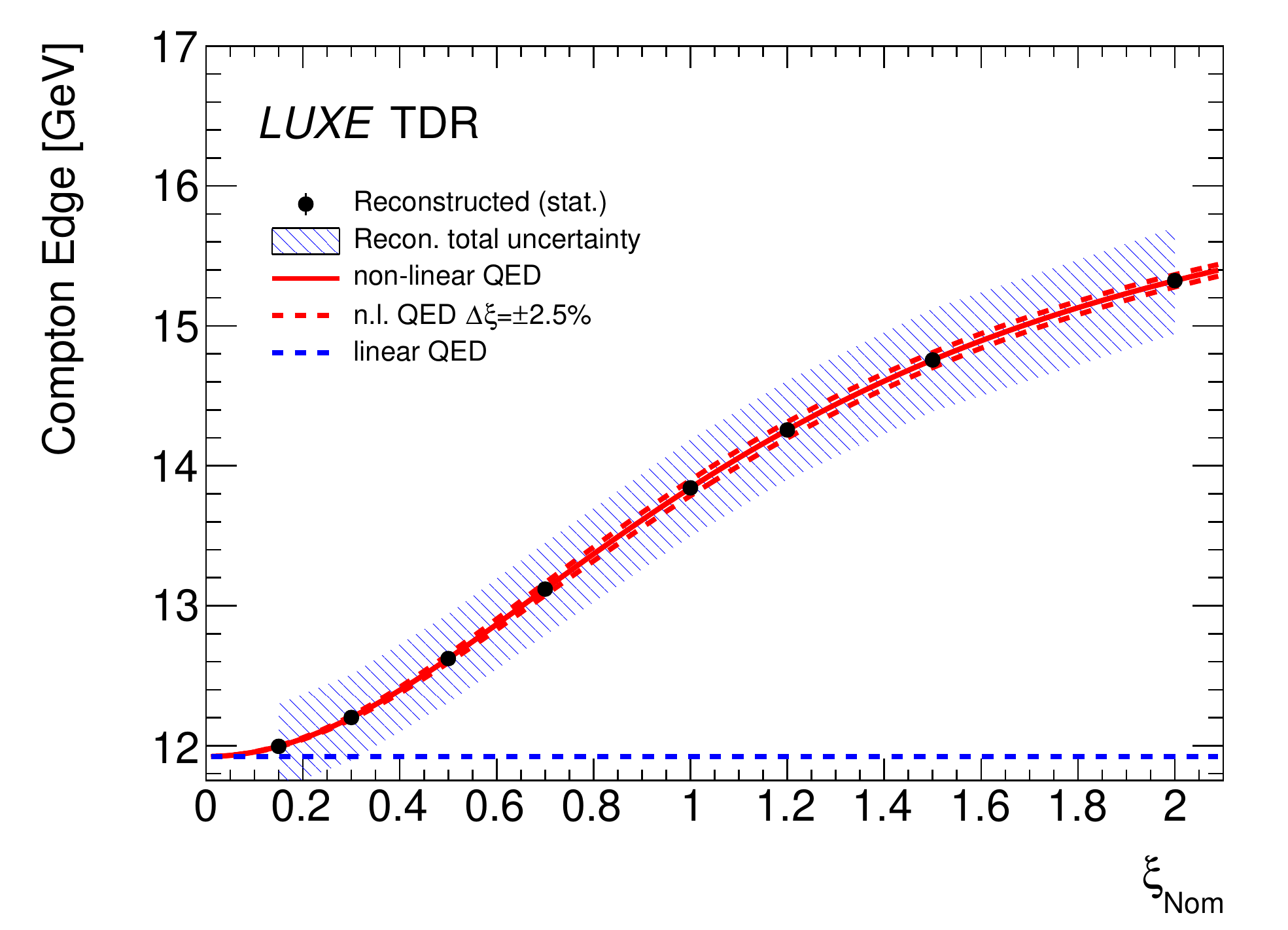}
         \caption{}
        \label{fig:comptonedge}
     \end{subfigure}
     \hfill
     \begin{subfigure}[b]{0.49\textwidth}
         \centering
         \includegraphics[width=\textwidth]{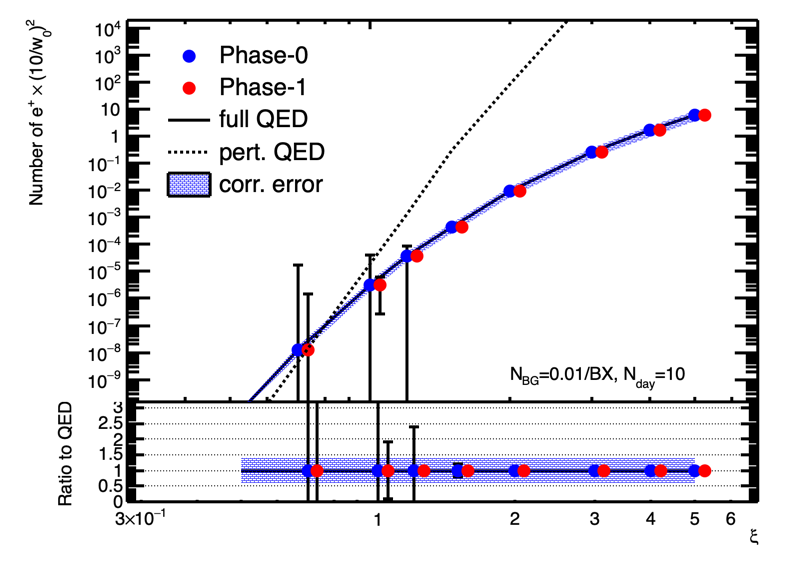}
        \caption{}    
         \label{fig:bpppprob}
     \end{subfigure}
      \caption{Left: Expected reconstructed Compton edge positions as a function of laser intensity, compared to linear and non-linear QED predictions. The band indicates a $2.5\%$ uncertainty on the energy scale. Statistical errors based on one hour of data-taking are shown but negligible. Right: Expected rate of Breit-Wheeler pair production per bunch crossing as a function of laser intensity in $\gamma$-laser collisions, assuming 10 days of data-taking and a background rate of $0.01$ event/bunch crossing.}
\end{figure}

Figure \ref{fig:comptonedge} shows the expected shift of the Compton edge position as a function of the laser intensity $\xi$ in $e^-$-laser collisions. Due to the large particle rate, already after one hour of data-taking the statistical uncertainties on the measured edge position are negligible. The shaded band shows the impact of a $2.5\%$ uncertainty on the energy scale, which is deemed achievable with the foreseen detectors. It is clearly visible that LUXE will be able to probe the shift of the Compton edge and establish the electron effective mass scaling $m_e\rightarrow m_e\sqrt{1+\xi^2}$ (solid red line) and the departure from the linear-QED prediction (dashed blue line). The impact of a $2.5\%$ uncertainty on the laser intensity measurement (dashed red lines) is small.\\

Figure \ref{fig:bpppprob} shows the expected rate of positrons created via the Breit-Wheeler pair production process in $\gamma$-laser collisions as a function of $\xi$. The statistical uncertainties after ten days of data-taking are small for $\xi>1.5$. Shown as the blue band is the impact of a $2.5\%$ uncertainty on the measured value of $\xi$, resulting in a $40\%$ uncertainty on the positron rate. This sizeable systematic uncertainty emphasizes the importance of high precision laser diagnostics in LUXE. The departure of the expected positron rate from the perturbative QED prediction is clearly visible.

\section{Conclusion}
\label{sec:conclusion}

LUXE aims to probe QED in a new regime of strong-fields, by studying collisions between an electron beam, or high-energy gamma photon beam, with a high-intensity optical laser. The design of LUXE as a collision experiment with continuous data-taking will enable precision measurements of strong-field QED processes, such as non-linear Compton scattering and Breit-Wheeler pair creation. The laser system and particle detectors in LUXE are custom-designed in order to meet their physics purpose. LUXE has a good chance to be the first to enter a regime of QED previously uncharted by experiments in clean laboratory conditions and will be the first experiment to study high-intensity laser collisions with real high-energy gamma photons.

\section*{Acknowledgements}
\addcontentsline{toc}{section}{Acknowledgements}
This work was in part funded by the Deutsche Forschungsgemeinschaft under Germany’s Excellence Strategy – EXC 2121 “Quantum Universe" – 390833306.

\printbibliography

\end{document}